\PassOptionsToPackage{hyphens}{url} 
\documentclass[%
 aip,
 amsmath,amssymb,
 reprint,%
]{revtex4-2}
\usepackage{dcolumn,mathptmx,setspace}
\usepackage{bbm}
\usepackage[colorinlistoftodos]{todonotes}
\usepackage[utf8]{inputenc} 
\usepackage[T1]{fontenc}    

\usepackage[utf8]{inputenc} 
\usepackage[noend]{algpseudocode}

\usepackage{amssymb,amsmath,mathrsfs,stmaryrd,amsthm,mathtools,amsfonts,graphicx,hyperref,url,booktabs,nicefrac,comment,microtype,tikz-cd,lingmacros,nccmath,tree-dvips, bbm, bm, etoolbox,algorithm,caption,subcaption,color,enumitem,relsize,soul}

\usepackage{amsmath}

\begin{document}

\preprint{AIP/123-QED} 
\title[In search of peak human athletic potential: A mathematical investigation]{In search of peak human athletic potential: A mathematical investigation}

\author{Nick James}
\affiliation{ 
School of Mathematics and Statistics, University of Melbourne, Victoria, 3010, Australia}%
\author{Max Menzies}
\email{max.menzies@alumni.harvard.edu}
\affiliation{%
Beijing Institute of Mathematical Sciences and Applications, Tsinghua University, Beijing, 101408, China}%
\author{Howard Bondell}
\affiliation{ 
School of Mathematics and Statistics, University of Melbourne, Victoria, 3010, Australia}%

\date{December 24 2021}
\begin{abstract}

This paper applies existing and new approaches to study trends in the performance of elite athletes over time. We study both track and field scores of men and women athletes on a yearly basis from 2001 to 2019, revealing several trends and findings. First, we perform a detailed regression study to reveal the existence of an ``Olympic effect'', where average performance improves during Olympic years. Next, we study the rate of change in athlete performance and fail to reject the notion that athlete scores are leveling off, at least among the top 100 annual scores. Third, we examine the relationship in performance trends among men and women's categories of the same event, revealing striking similarity, together with some anomalous events. Finally, we analyze the geographic composition of the world's top athletes, attempting to understand how the diversity by country and continent varies over time across events. We challenge a widely held conception of athletics, that certain events are more geographically dominated than others. Our methods and findings could be applied more generally to identify evolutionary dynamics in group performance and highlight spatio-temporal trends in group composition.

\end{abstract}

\maketitle

\begin{quotation}

Throughout time, humans have aspired to improve in a wide range of societal aspects. Over the last several hundred years, we have seen significant advances in technology, health, the economy, education, and other fields. Further progress in such areas of human life is expected to continue in the coming years. In areas such as sports and athletics, there may be an ultimate limit to human potential. The recent completion of The Tokyo 2020 Olympic and Paralympic Games piqued our interest in studying evolutionary patterns in human athletic performance over the last two decades. Throughout this paper, we seek to better identify patterns and structural similarities in men and women's track and field events. We seek to identify anomalous trajectories, explore the existence and potential impact of an ``Olympic effect'' on collective athlete performance, measure the similarity of performance trends between men and women, and study the variance in geographic representation of events' top athletes. We hope that this work will further encourage the application of techniques from the nonlinear dynamics community to better understand time-varying behaviors in athletics, and sports more generally.

\end{quotation}

\section{Introduction}

Every year, there are many competitions at which the world's top athletes participate. Athletic sports comprise two main categories: track, in which athletes are scored by their times (and a lower measured time is a better result), and field, in which athletes are scored by a distance thrown or jumped (and a higher distance is a better result). Typically, competitions consist of both track and field events. In all events, elite athletes seek the marginal improvements necessary to best their opponents, where seconds, split seconds, and centimeters may make the difference. Among all competitions, the Olympic Games, held every four years, are the most well-known and prestigious, attracting competitors worldwide. The modern Olympic Games were first held in 1896 in Athens, Greece, and winning an Olympic gold medal is considered one of sport's greatest achievements.

Given the intense competition around score improvements, athletes, commentators, analysts, and fans alike all share an interest in closely tracking the best results over time. There are numerous reasons for understanding the trends in these results, including fan interest, the identification of avenues for improvement, certain countries performing better than others, allocation of dedicated resources to burgeoning young athletes, and identifying athletes with promising growth trajectories who could become suitable brand ambassadors for various companies.

For this purpose, there has been a substantial body of statistical research on sports conducted both on a private and academic basis.\cite{Elderton1945,Kimber1993,Gabel2012,Pawlowski2013,Wu2021_sport} In particular, statisticians have applied extreme value theory to investigate records, namely singularly outstanding performances, in athletics.\cite{Einmahl2008,Robinson1995,Gembris2002,Baro1999} However, there have been limited applications of nonlinear dynamics and physically-inspired mathematics to understand patterns and trends across various sports.\cite{Ribeiro2012,Merritt2014,Clauset2015,Kiley2016} Our paper builds on a long literature of \emph{multivariate time series analysis}, methods that are more frequently applied to other domains, including epidemiology,\cite{james2021_CovidIndia,james2020covidusa,Chowell2016,jamescovideu,Manchein2020,Blasius2020,James2021_virulence,james2021_TVO} finance,\cite{Drod2020_entropy,Jamesfincovid,Drod2021_entropy,james2021_mobility,Drod2020,james2021_MJW,Wtorek2020,arjun} and other fields.\cite{Vazquez2006,Mendes2018,Shang2020} Methods of time series analysis are broad, including parametric models, \cite{Hethcote2000,Perc2020} distance analysis and correlation, \cite{Moeckel1997,Szkely2007,Mendes2019,James2020_nsm}  network models, \cite{Karaivanov2020,Ge2020,Xue2020} clustering \cite{Machado2020} and many others. \cite{Ngonghala2020,Cavataio2021,Nraigh2020,Glass2020} In this paper, we draw upon numerous existing techniques, including linear regression,\cite{Hastie2009} trajectory analysis \cite{NguenaNguefack2020, Nagin2014}, inconsistency analysis,\cite{James2021_crypto,james2021_crypto2} agglomerative hierarchical clustering \cite{Mllner2013} and geographic dispersion.\cite{James2021_geodesicWasserstein}

This paper aims to study trends and relationships between the performance (and composition) of male and female athletes.  First, Section \ref{sec:Olympic effect} performs several regressions to investigate the existence and extent of an ``Olympic effect,'' in which athletes' scores are usually (but not always) superior during Olympic years. Section \ref{sec:curvature} examines the linearity of score increments more closely to investigate whether there is evidence of a recent leveling off in average scores. Subsequently, Section \ref{sec:AlignmentAnalysis} takes a closer analysis of the yearly changes in scores and shows broad alignment between male and female performance in corresponding events. Next, we analyze the trajectories of performances over time and determine inconsistencies between the performance of men and women in Section \ref{sec:TrajectoryModelling}. Finally, Section \ref{sec:GeographicConcentration} analyzes the composition of top athletes on an event-by-event basis. We track the extent of the geographic spread of the best athletes across the world over time. In particular, we show surprising uniformity in this variance, challenging the notion that some events are more dominated by smaller collections of countries than others.

\section{Data}

Our data consists of the top 100 scores (times and distances, respectively) recorded each year in the men and women's competitions of 16 track and field events. Our data source is World Athletics (\url{https://www.worldathletics.org}). Due to the cancellation of many sporting events during COVID-19 and 2021's being incomplete, we exclude 2020 and 2021 from our analysis window. Thus, our data spans $t=2001,...,2019.$ Track events are scored in times, with shorter times scoring better, while field events are scored in distances, with larger distances scoring better. Thus, we consistently analyze track and field events separately, and briefly consider the difference in behavior between the top $m=10$ and 100 scores in each event. Our events are the following: high jump, long jump, pole vault, triple jump, discus, hammer throw, javelin, shot put, 10 000 meters (10K), 5000 meters (5K), 3000 meters (3K), 1500 meters (1500m), 800m, 400m, 200m and 100m. All these running events are of a flat elevation. Throughout the manuscript, let $x_i(t), y_i(t)$ be the average of the top $m$ scores in the women and men's competition of the event indexed by $i$.

\begin{table*}
\begin{center}
\begin{tabular}{ |p{3.5cm}||p{2.3cm}|p{2.3cm}|p{2.3cm}|p{1.65cm}|p{1.6cm}|p{1.6cm}|p{1.6cm}|}
 \hline
 \multicolumn{8}{|c|}{Comparison of linear regression models} \\
 \hline
 Event & Model 1 Adj. $R^2$ & Model 2 Adj. $R^2$ & Model 3 Adj. $R^2$ & $\beta_1$ Top 100 & $\alpha_0$ Top 100 & $\beta_1$ Top 10 & $\alpha_0$ Top 10 \\
 \hline
 Men's high jump & 0.24 & 0.33 & 0.26 & .0005$^{**}$ & .0046$^{*}$ & .0009$^{**}$ & .007 \\
 Women's high jump & 0.13 & 0.24 & 0.45 & -.0004$^{**}$ & .005$^{*}$ & -.0008$^{*}$ & .0073  \\
 Men's long jump & 0.11 & 0.16 & 0.13 & .0014$^{*}$ & .016 & .003$^{*}$ & -.01 \\  
 Women's long jump & 0.26 & 0.47 & 0.61 & .003$^{*}$ & .036$^{**}$ & .0021 & .073$^{**}$ \\  
 Men's pole vault & -0.03 & 0.0014 & -0.05 & .0007 & .016 & .0006 & -.004 \\  
 Women's pole vault & 0.89 & 0.93 & 0.94 & 0.01$^{***}$ & 0.035$^{**}$ & .01$^{***}$ & .04$^{**}$ \\  
 Men's triple jump & -0.06 & 0.13 & 0.08 & -.0002 & .066$^{**}$ & -.0003 & -.02 \\  
 Women's triple jump & 0.06 & 0.30 & 0.34 & -.0042 & .085$^{**}$ & -.01$^{**}$ & .16$^{***}$ \\  
 Men's discus & 0.27 & 0.71 & 0.68 & .039$^{***}$ & .62$^{***}$ & -.04$^{*}$ & .33 \\  
 Women's discus & 0.04 & 0.52 & 0.50 & .024$^{*}$ & .75$^{***}$ & .06$^{**}$ & .37 \\  
 Men's hammer throw & 0.54 & 0.79 & 0.84 & -.08$^{***}$ & .67$^{***}$ & -.21$^{***}$ & .4 \\  
 Women's hammer throw & 0.60 & 0.71 & 0.70 & 0.19$^{***}$ & 1.16$^{**}$ & .23$^{***}$ & .63 \\  
 Men's javelin & 0.49 & 0.51 & 0.46 & 0.09$^{***}$ & 0.33 & .05 & -.67 \\  
 Women's javelin & 0.63 & 0.73 & 0.70 & 0.08$^{***}$ & 0.45$^{**}$ & .09$^{***}$ & .24 \\  
 Men's shot put & 0.71 & 0.73 & 0.70 & 0.033$^{***}$ & 0.095 & .04$^{***}$ & -.08 \\  
 Women's shot put & 0.05 & 0.24 & 0.15 & .007 & .138$^{**}$ & -.01$^{**}$ & .24$^{***}$ \\  
  Men's 10K & 0.12 & 0.37 & 0.66 & -.58$^{**}$ & -9.9$^{**}$ & 0.27 & -5.8 \\  
 Women's 10K & -0.05 & 0.17 & 0.29 & -.2 & -13.7$^{**}$ & .11 & -26.8$^{**} $\\  
 Men's 5K & -0.05 & 0.03 & -0.09 & 0.04 & -1.9 & .19 & -.92 \\  
 Women's 5K & 0.003 & 0.16 & 0.14 & -.17 & -4.0$^{*}$ & -.46$^{***}$ & -1.3 \\ 
 Men's 3K & 0.50 & 0.53 & 0.57 & 0.3$^{***}$ & 1.3 & .14 & -.33 \\  
 Women's 3K & -0.06 & -0.07 & 0.15 & -.0006 & 1.7 & -.16 & 4.0 \\ 
  Men's 1500m & 0.06 & 0.08 & 0.19 & -.023 & -.25 & .02 & .26 \\  
 Women's 1500m & 0.69 & 0.70 & 0.66 & -.13$^{***}$ & -.36 & -.16$^{***}$ & .05 \\ 
 Men's 800m & 0.54 & 0.63 & 0.60 & -.03$^{***}$ & -.15$^{**}$ & -.02 & .24 \\  
 Women's 800m & 0.20 & 0.58 & 0.53 & -0.02$^{**}$ & -0.4$^{***}$ & -.008 & -.63$^{**}$ \\ 
 Men's 400m & 0.58 & 0.55 & 0.52 & -0.2$^{***}$ & -.02 & -.03$^{***}$ & -.05 \\  
 Women's 400m & 0.05 & 0.30 & 0.23 & -.008 & -.18$^{**}$ & -.01 & -.12 \\ 
 Men's 200m & 0.86 & 0.86 & 0.85 & -.014$^{***}$ & -.015 & -.01$^{***}$ & -.02 \\  
 Women's 200m & 0.73 & 0.86 & 0.85 & -.016$^{***}$ & -.09$^{***}$ & -.02$^{***}$ & -.11$^{**}$ \\ 
 Men's 100m & 0.85 & 0.89 & 0.89 & -.007$^{***}$ & -.02$^{**}$ & -.005$^{***}$ & -.04$^{*}$ \\  
 Women's 100m & 0.87 & 0.89 & 0.90 & -.009$^{***}$ & -.02$^{**}$ & -.009$^{***}$ & -.05$^{**}$ \\ 
\hline
\end{tabular}
\caption{First, we record the adjusted $R^2$ values for three linear models applied to the top $m=100$ scores and observe that model 2 is generally the best fit. Next, we record regression coefficients for model 2, both for $m=100$ and $m=10$. We highlight $\beta_1$ and $\alpha_0$, the year and Olympic indicator coefficients, respectively. We indicate associated $p$-values of $p<0.01, p<0.05, p<0.1$ with $^{***}$, $^{**}$ and $^{*}$, respectively. }
\label{tab:regressioncoefficients}
\end{center}
\end{table*}

\section{Olympic effect regressions}
\label{sec:Olympic effect}

In this section, we test the existence of an \emph{Olympic effect} - a phenomenon where athletic performance is superior during Olympic years. To study this effect, we fit several linear regression models for both men and women's events. In model 1, we assume that athletic performance is a linear function over time subject to additive white noise. For each men's event $i$, model 1 is formulated as follows:
\begin{align}
\label{eq:model1}
  y_i(t) = \beta_0 + \beta_1 (t-2000) + \epsilon(t).
\end{align}
Here, $y_i(t)$ is the mean of the top $m$ scores in year $t$, $\beta_0$ is the model's intercept and $\beta_1$ is the trend of event performance over time, both of which we estimate in the modeling procedure. We implement this for both $m=10$ and 100, averaging the top 10 and 100 scores per year, respectively. As is typical in an ordinary least squares (OLS) framework, $\epsilon(t)$ is assumed to be independently and identically distributed from a Gaussian distribution centered at zero. In Appendix \ref{app:regression}, we include some necessary validation of technical hypotheses that underpin the reliability of the linear regression model.

In model 2, we assume that athletic performance is a linear function over time and explicitly model an Olympic effect with an indicator function during Olympic years. Model 2 is formulated as 
\begin{align}
\label{eq:model2}
  y_i(t) = \beta_0 + \beta_1 (t-2000) + \alpha_0  \mathbbm{1}_{ \mathbb{O} } (t) + \epsilon(t).
\end{align}
Here, $\mathbbm{1}_{ \mathbb{O}}$ is an indicator function taking a value of 1 only during Olympic years, so $\alpha_0$ measures the extent of this effect. 

Finally, model 3 explores the possibility of 4-year periodic behavior in scores not primarily driven by Olympic years. Specifically, we fit the response variable $y_i(t)$ of mean scores against not only year $t$, but also a categorical variable according to whether $t$ is 0,1,2 or 3 (modulo 4). We encode this with three dummy variables as follows:
\begin{align}
  y_i(t) = \beta_0 + \beta_1 (t-2000) 
  + \alpha_1 \mathbbm{1}_1 (t) +\alpha_2 \mathbbm{1}_2 (t)+\alpha_3 \mathbbm{1}_3 (t)+ \epsilon(t).
\end{align}
Here, $\mathbbm{1}_j (t)$ is an indicator function taking value 1 if $t \equiv j$ (mod 4). Then $\alpha_1,\alpha_2$ and $\alpha_3$ are the linear coefficients for years that are 1,2 and 3 modulo 4, respectively. As is standard with dummy encoding of categorical variables, four separate categories are encoded with three numeric indicator variables. It is arbitrary which category is not included in this encoding, so we have chosen to distinguish the coefficient of Olympic years by not including an $\alpha_0$ term here. All three models are written analogously for women's events, using $x_i(t)$ as our response variable. 

We fit these three models to all track and field events for both men and women and explore the insights therein. Our general approach is to find a simple model that generalizes well across most events. Accordingly, we validate our models' estimates using adjusted $R^2$, where additional parameters yield a complexity penalty. Table \ref{tab:regressioncoefficients} shows the adjusted $R^2$ for all three models, where one can see that model 2 is generally the best candidate model. Specifically, model 2 has the best adjusted $R^2$ score in 58\% of events.

In the final four columns of the table, we present $\alpha_0$ and $\beta_1$ for model 2, in the cases where $m=100$ and $m=10$ respectively. $\beta_1$ represents the coefficient of the linear term, and $\alpha_0$ represents the coefficient of the Olympic indicator function. Table \ref{tab:regressioncoefficients} demonstrates that most events display an improving linear trend in performance, with a positive Olympic effect (greater distances for field events and lower times for track events). In the case where $m=100$, all field events exhibit a positive $\alpha_0$ term, with the majority being statistically significant, an Olympic effect where performance improves during Olympic years. Findings on the track are similar, where all but two events (the men and women's 3000 meters) display a negative $\alpha_0$ term, indicating improved track performance during Olympic years.

Interestingly, this Olympic effect appears to be less ubiquitous in the case where $m=10$. Five field events (men's long jump, men's pole vault, men's triple jump, men's javelin, and men's shot put) display a negative Olympic effect, while those that are positive are mostly not statistically significant. That is, among the top 10 athletes, we do not observe evidence that Olympic years are associated with increased average athletic performance. Similarly, with track, Olympic years provide less consistently positive boosts in performance. Four track events (men's 800m, women's 1500m, men's 1500m, and women's 3000m) have a positive (but not significant) $\alpha_0$ term, again indicating less evidence for improved performance during Olympic years.

\begin{figure*}
    \centering
    \begin{subfigure}[b]{0.49\textwidth}
        \includegraphics[width=\textwidth]{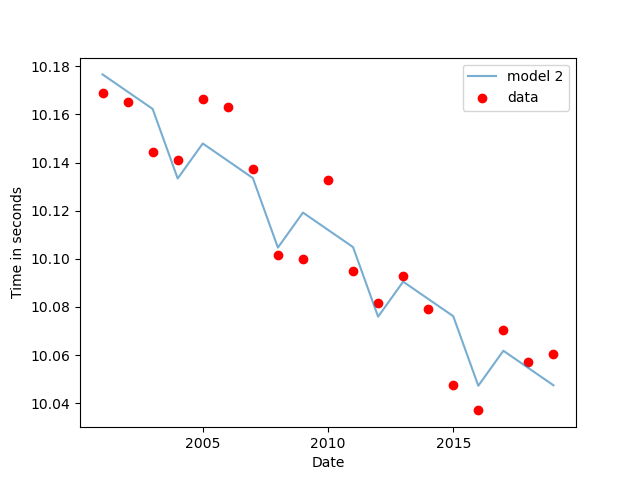}
        \caption{}
        \label{fig:100m_men}
    \end{subfigure}
    \begin{subfigure}[b]{0.49\textwidth}
        \includegraphics[width=\textwidth]{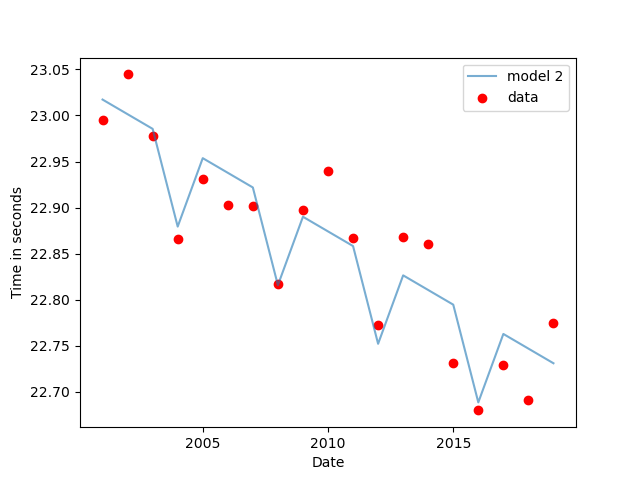}
        \caption{}
        \label{fig:200m_women}
    \end{subfigure}
\begin{subfigure}[b]{0.49\textwidth}
        \includegraphics[width=\textwidth]{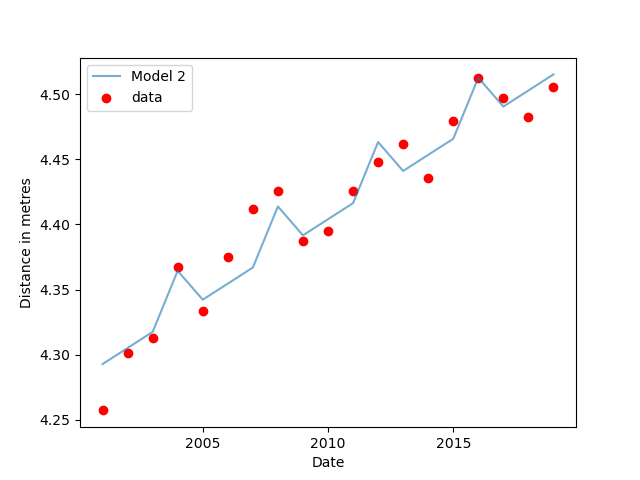}
        \caption{}
        \label{fig:polevault_women}
    \end{subfigure}
\begin{subfigure}[b]{0.49\textwidth}
        \includegraphics[width=\textwidth]{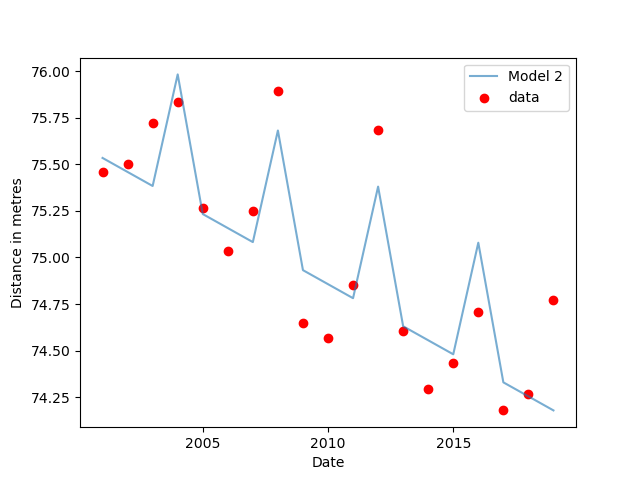}
        \caption{}
        \label{fig:hammerthrow_men}
    \end{subfigure}
    \caption{Model 2 regression fits for the top $m=100$ athletes in (a) the mens' 100 meters, (b) women's 200 meters, (c) women's pole vault, (d) men's hammer throw. All events exhibit a linear trend with a pronounced Olympic effect. All except the hammer throw exhibit better performance with time (lower times for track and higher distances for field). The hammer throw is an anomaly, exhibiting worse performance with time.}
   \label{fig:OlympicEffectRegressionModels}
\end{figure*}

For most track and field events, especially with $m=100$, model 2 does a good job of capturing the true essence of the data: a linear trend accompanied by a (generally positive) effect on average performance during Olympic years. Figure \ref{fig:OlympicEffectRegressionModels} shows model 2's estimates for four events: two track events and two field events when $m=100$. Figures \ref{fig:100m_men} and \ref{fig:200m_women}, which display the men's 100m and women's 200m, respectively, show typical results for track events. Each features an evident linear trend of reduced times and Olympic effect of improvement during Olympic years.

Turning to the field events, Figure \ref{fig:polevault_women}, displaying the women's pole vault, shows a typical result for the field. A positive linear term is evident, highlighting consistent improvement in average pole vault performance between 2001 and 2019, as well as spikes of improved performance during Olympic years. Finally, Figure \ref{fig:hammerthrow_men} highlights an anomalous trend in the men's hammer throw. Surprisingly, average performance is observed to decline with time, although a beneficial Olympic effect persists.


\section{First differences regression}
\label{sec:curvature}

In this brief section, we are interested in whether we can observe whether athletic scores are approaching a peak, slowing down in incremental improvements. We now examine this more closely by analyzing the first differences in scores $\Delta x_i (t) = x_i(t+1) - x_i(t)$, $i=1,...,N, t=2001,...,2018$, similarly for $\Delta y_i$. In this section, we fix $m=100$, always considering the average of the top 100 scores each year. Unlike other literature that studies the evolution of world records or the best result each year, this paper seeks to determine trends in overall athlete performance. Accordingly, we do not run our analysis when $m=1$, as the underlying trends in performance would be subject to outliers, and may become too volatile. This becomes even more severe when we look at first differences (effectively a first derivative), which is always more irregular than the original data. If we were to study trends in the men's 100 metres where $m=1$, one would identify peak performance in 2009, and since then, a decline in performance. However, this peak corresponds to Usain Bolt's exceptional time of 9.58 seconds.\cite{olympicsdata} To put this result in perspective, the three best times in the men's 100 metres in the 2019 season were 9.76 seconds, 9.86 seconds and 9.87 seconds \cite{olympicsdata} - notably behind Bolt's world record time. One can see how the study of too few samples within each year may lead to erroneous insights as to whether collective performance of elite athletes in various events is improving or declining. Echoing Section \ref{sec:Olympic effect}, we formulate two models for these differences. Model 1 is formulated as follows:
\begin{align}
  \Delta y_i(t) = \beta_0 + \beta_1 (t-2000) + \epsilon(t),
\end{align}
analogously to (\ref{eq:model1}). It aims to track the differences in improvement relative to time. Model 2 is formulated analogously to (\ref{eq:model2}),
\begin{align}
  \Delta y_i(t) = \beta_0 + \beta_1 (t-2000)+ \alpha_0  (\mathbbm{1}_{ \mathbb{O} } (t+1) -  \mathbbm{1}_{ \mathbb{O} }(t)) + \epsilon(t),
\end{align}
including a difference term that incorporates the predicted increase during and decrease after an Olympic year. Table \ref{tab:FirstDifferencesCoefficients} displays the determined $\beta_1$ coefficients for each model, and the $\alpha_0$ term from model 2. Importantly, not a single associated $p$-value associated with $\beta_1$ in either model is less than 0.1, offering no evidence from either model in any event that the rate of incremental improvements is changing over time. On the other hand, even in the first difference model, most values of $\alpha_0$ are positive and significant for field events, and negative and significant for track events. That is, from both Section \ref{sec:Olympic effect} and \ref{sec:curvature}, we observe linear fits in improving performance, with clear evidence of an Olympic effect, but no evidence of diminishing returns or nearing the natural peak of human performance, at least in the mean of the top 100 scorers over this period.

\begin{table*}
\begin{center}
\begin{tabular}{ |p{3.5cm}||p{2cm}|p{2cm}|p{3cm}|}
 \hline
 \multicolumn{4}{|c|}{Trends in first differences} \\
 \hline
 Event & Model 1 $\beta_1$ & Model 2 $\beta_1$ & Model 2 $\alpha_0$\\
 \hline
 Men's high jump & -3.24 x 10$^{-4}$ & -3.05 x 10$^{-4}$ & 2.25 x 10$^{-3}$ \\
 Women's high jump & -6.32 x 10$^{-5}$ & -4.01 x 10$^{-5}$ &  2.27 x 10$^{-3}$\\
 Men's long jump & -2.90 x 10$^{-4}$ & 1.92 x 10$^{-4}$ & 1.20 x 10$^{-2}$\\  
 Women's long jump & -5.86 x 10$^{-4}$ & 3.64 x 10$^{-4}$ & 2.68 x 10$^{-2}$ $^{**}$\\  
 Men's pole vault & 6.02 x 10$^{-4}$ & 7.31 x 10$^{-4}$ & 1.56 x 10$^{-2}$  $^{*}$\\  
 Women's pole vault & -1.17 x 10$^{-3}$ & 9.71 x 10$^{-4}$ &  2.44 x 10$^{-2}$  $^{**}$\\  
 Men's triple jump & -1.75 x 10$^{-4}$ & 3.20 x 10$^{-4}$ & 5.99 x 10$^{-2}$  $^{***}$\\  
 Women's triple jump & -2.02 x 10$^{-3}$ & -1.44 x 10$^{-3}$& 6.99 x 10$^{-2}$  $^{***}$\\  
 Men's discus & -3.07 x 10$^{-3}$ & 2.40 x 10$^{-3}$ & 6.62 x 10$^{-1}$  $^{***}$\\  
 Women's discus & 6.30 x 10$^{-3}$ & 1.25 x 10$^{-2}$ & 7.50 x 10$^{-1}$  $^{***}$\\  
 Men's hammer throw & 4.57 x 10$^{-3}$ & 1.01 x 10$^{-2}$ & 6.67 x 10$^{-1}$  $^{***}$\\
 Women's hammer throw & -4.35 x 10$^{-2}$ & -3.70 x 10$^{-2}$ & 7.50 x 10$^{-1}$  $^{***}$\\  
 Men's javelin & 1.71 x 10$^{-2}$ & 2.04 x 10$^{-2}$ & 3.94 x 10$^{-1}$  $^{**}$\\  
 Women's javelin & 3.28 x 10$^{-3}$ & 7.10 x 10$^{-3}$ & 4.62 x 10$^{-1}$  $^{***}$\\  
 Men's shot put & 4.72 x 10$^{-3}$ & 5.78 x 10$^{-3}$ & 1.29 x 10$^{-1}$  $^{***}$\\  
 Women's shot put & 3.75 x $10^{-3}$ & 5.05 x 10$^{-3}$ & 1.57 x 10$^{-1}$  $^{***}$\\  
  Men's 10K & 1.8 x 10$^{-1}$ & 1.03 x 10$^{-1}$ & -6.34  $^{*}$ \\  
 Women's 10K & -2.38 x 10$^{-1}$ & -3.33 x 10$^{-1}$ & -11.57   $^{**}$\\  
 Men's 5K & 5.85 x 10$^{-2}$  & 4.60 x 10$^{-2}$ & -1.43  $^{**}$ \\  
 Women's 5K & -1.24 x 10$^{-1}$ & -1.3 x 10$^{-1}$ & -4.24  $^{***}$\\  
 Men's 3K & 4.99 x 10$^{-2}$ & 6.32 x 10$^{-2}$ & 1.993  $^{***}$ \\  
 Women's 3K & -9.22 x 10$^{-2}$ & -1.04 x 10$^{-1}$  & -4.67 x 10$^{-1}$ \\  
  Men's 1500m & -3.43 x 10$^{-3}$ & -3.75 x 10$^{-3}$ & -1.52 x 10$^{-1}$ \\  
 Women's 1500m & -5.7 x 10$^{-3}$ & -9.21 x 10$^{-3}$ & -3.98 x 10$^{-1}$  $^{**}$\\  
 Men's 800m & -2.62 x 10$^{-3}$ & -3.85 x 10$^{-3}$ & -1.49 x 10$^{-1}$  $^{**}$\\  
 Women's 800m & 1.57 x 10$^{-2}$ & -9.21 x 10$^{-3}$ & -3.29 x 10$^{-1}$  $^{***}$\\  
 Men's 400m & -1.16 x 10$^{-3}$ & -1.40 x 10$^{-3}$ & -2.89 x 10$^{-2}$ \\  
 Women's 400m & 1.71 x 10$^{-3}$ & 4.34 x 10$^{-4}$ & -1.55 x 10$^{-1}$  $^{***}$ \\  
 Men's 200m & -5.43 x 10$^{-4}$ & -6.60 x 10$^{-4}$ & -1.41 x 10$^{-2}$  \\  
 Women's 200m & 1.16 x 10$^{-3}$ & 5.19 x 10$^{-4}$ & -7.87 x 10$^{-2}$  $^{***}$\\  
 Men's 100m & 2.84 x 10$^{-4}$ & 1.49 x 10$^{-4}$ & -1.64 x 10$^{-2}$  $^{**}$\\  
 Women's 100m & 1.9 x 10$^{-4}$ & 5.30 x 10$^{-5}$ & -1.71 x 10$^{-2}$  $^{*}$\\  
\hline
\end{tabular}
\caption{Regression coefficients $\beta_1$ and $\alpha_0$ for the year and Olympic indicator, respectively, for two models that track the first differences in scores over time. We indicate associated $p$-values of $p<0.01, p<0.05, p<0.1$ with $^{***}$, $^{**}$ and $^{*}$, respectively. Broad similarity is observed between each model's $\beta_1$ coefficients, with not a single $\beta_1$ determined to be significant. Thus, our two models produce no evidence of concavity in the trend of the top 100 scores.}
\label{tab:FirstDifferencesCoefficients}
\end{center}
\end{table*}

\section{Alignment analysis and clustering}
\label{sec:AlignmentAnalysis}

In this section, we further analyze the first difference in scores $\Delta x_i, \Delta y_i$. Individual examinations of the data suggest a substantial similarity in the year-on-year changes between men and women's performances within the same event. That is, when the mean score in the men's exhibits an increase or decrease, the same is usually seen in the womens' scores for the same event. To investigate this quantitatively, we compute the normalized inner product $\eta_i=<\Delta x_i, \Delta y_i>_n$ for each event $i$:
\begin{align}
\label{eq:alignment}
\eta_i = \frac{\sum_{t=2001}^{2018} \Delta x_i(t) \Delta y_i(t)   }{ \left(\sum_{t=2001}^{2018} \Delta x_i(t)^2 \right)^\frac12 \left(\sum_{t=2001}^{2018} \Delta y_i(t)^2 \right)^\frac12  }    
\end{align}
This computes the \emph{alignment} between the first differences for each event between the yearly changes in men and women's scores. As we use a normalized inner product, it is appropriate to compare different time series with different scales (including different growth rates). More broadly, we compute a $2N \times 2N$ matrix $M$ defined by
\begin{align}
\label{eq:alignmentmatrix}
M_{ij}=\begin{cases}
<\Delta x_i, \Delta x_j>_n\text{ if } 1\leq i,j \leq N,\\
<\Delta y_{i-N},\Delta y_{j-N}>_n \text{ if } N+1\leq i,j \leq 2N,\\
<\Delta x_i, \Delta y_{j-N}>_n \text{ if } 1\leq i \leq N<j \leq 2N.
\end{cases}
\end{align}
This matrix computes normalized inner products between all first difference sequences for both men and women's events. Indices $i=1,...,N$ correspond to female first difference time series $\Delta x_i$, while $i=N+1,...,2N$ correspond to male first difference time series $\Delta y_i$. This matrix is constructed to compute inner products between all events of each gender. The subscript $n$ refers to the fact that the inner product is normalized, while $N=16$ is the total number of events. All these values are normalized to lie in $[-1,1]$.

\begin{figure}
    \centering
    \includegraphics[width=0.49\textwidth]{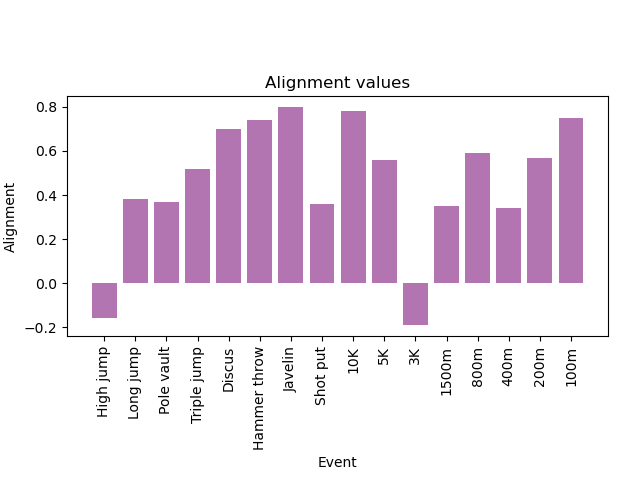}
    \caption{Alignment values $\eta_i$, defined in (\ref{eq:alignment}), between men and women's categories of the same sport. Broadly high values indicate significant alignment between year-on-year changes in men and women's scores.}
    \label{fig:Alignment}
\end{figure}

First, we display the alignment values $\eta_i=<\Delta x_i, \Delta y_i>$ for all events in Figure \ref{fig:Alignment}. Across all events, the median value is 0.54, which is relatively high given that all values lie in $[-1,1]$ and under no association would be centered around zero. The median alignment of field events is slightly lower than that of track events, from 0.45 to 0.57 respectively. The javelin has the greatest alignment between men and women's first differences with an alignment of 0.8, while the 3000m has the lowest alignment of -0.19.

\begin{figure*}
    \centering
    \begin{subfigure}[b]{0.75\textwidth}
        \includegraphics[width=\textwidth]{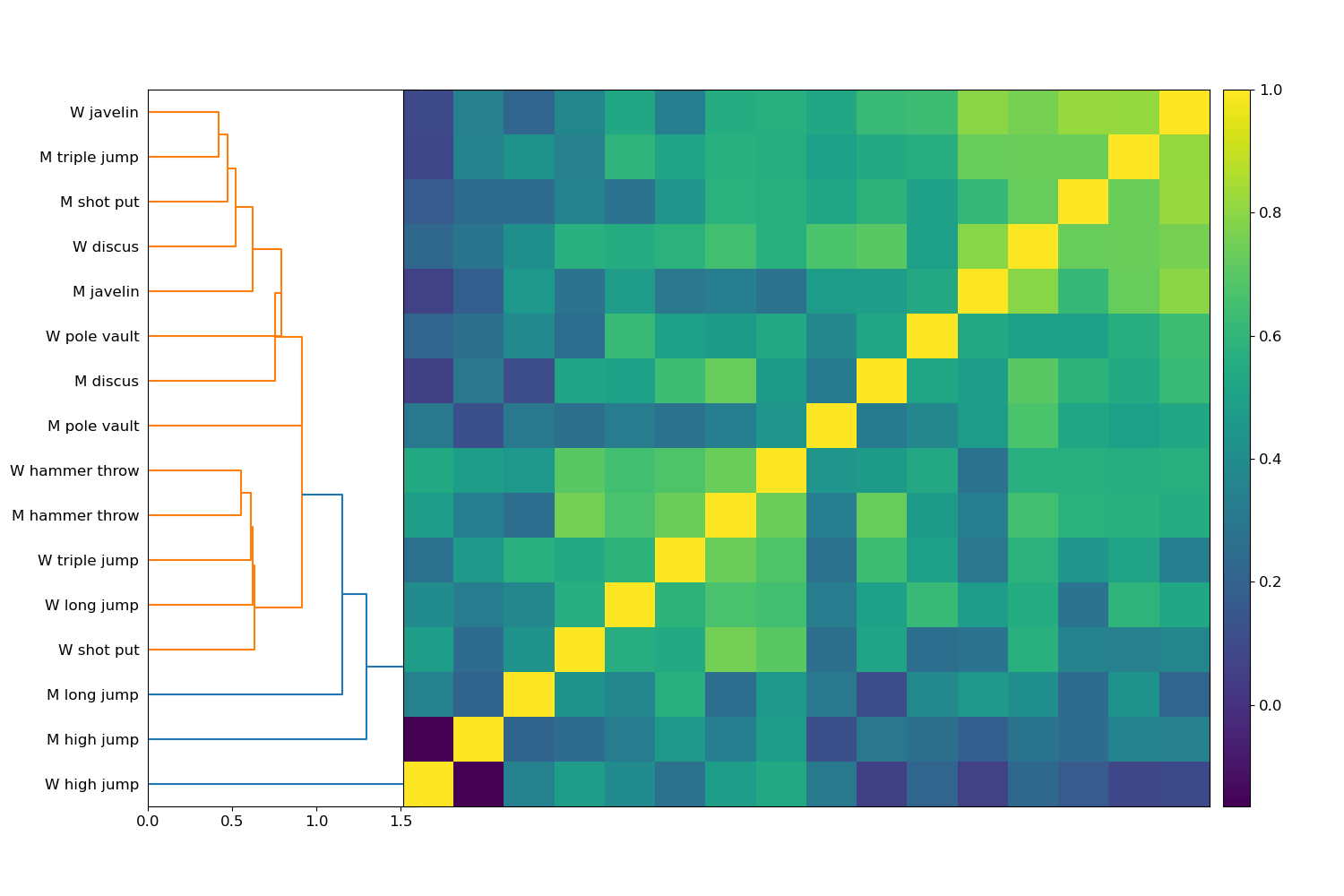}
        \caption{}
        \label{fig:Field_dendrogram}
    \end{subfigure}
    \begin{subfigure}[b]{0.75\textwidth}
        \includegraphics[width=\textwidth]{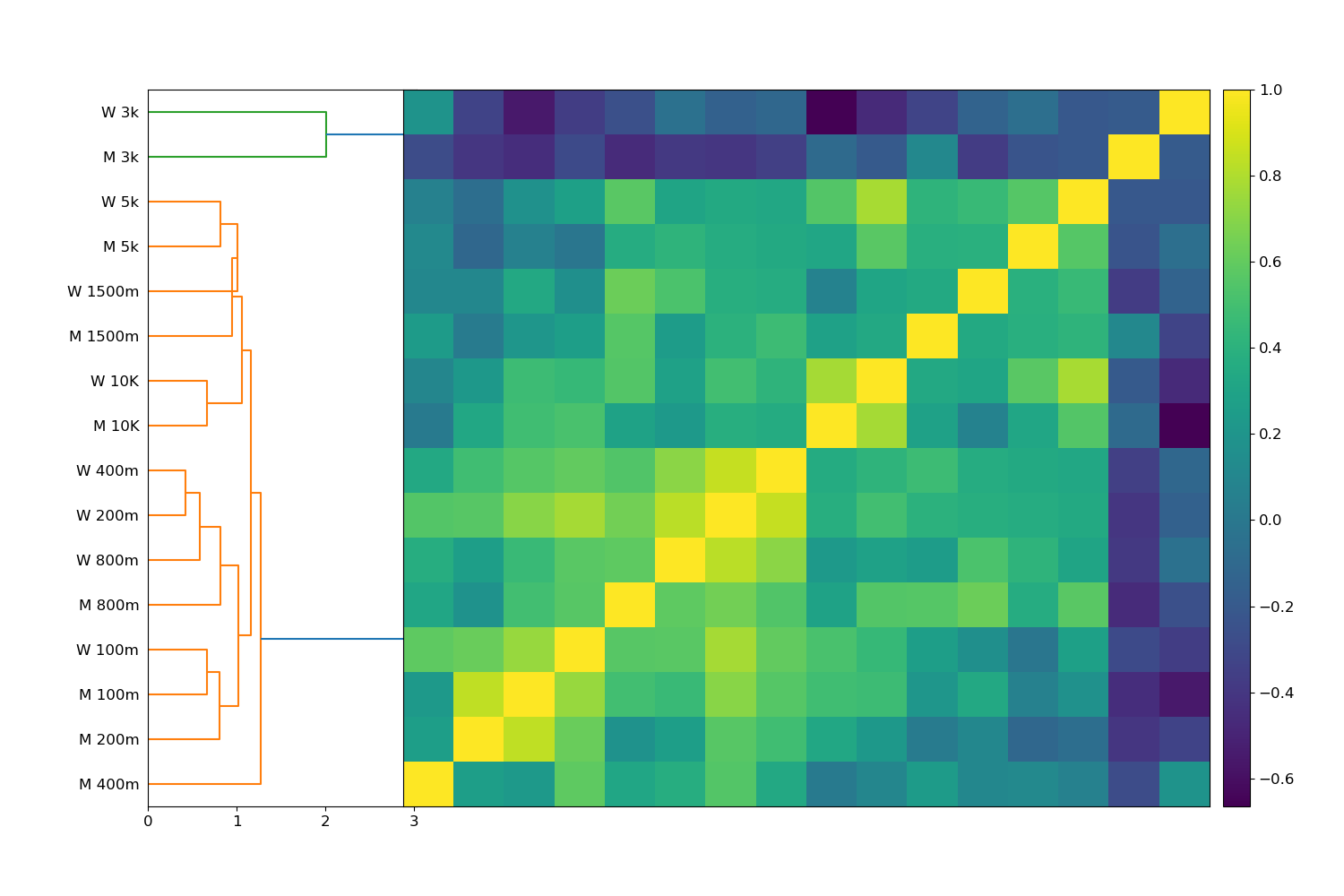}
        \caption{}
        \label{fig:Track_dendrogram}
    \end{subfigure}
    \caption{Hierarchical clustering on the matrix $M$, defined in (\ref{eq:alignmentmatrix}), measuring the extent of alignment in first differences between events. Numerous men and women's categories of the same event cluster together, both for (a) field and (b) track events. In addition, clustering is also observed between similar events, potentially due to intersection in the athletes performing in them.}
    \label{fig:AlignmentDendrograms}
\end{figure*}

To further explore alignment pairings between all candidate events, we perform hierarchical clustering on $M$ in Figure \ref{fig:AlignmentDendrograms}. Specifically, we implement agglomerative hierarchical clustering with the average-linkage criterion.\cite{Mllner2013} Figures \ref{fig:Track_dendrogram} and \ref{fig:Field_dendrogram} display the dendrograms of track and field events, respectively. Both dendrograms display a similar theme, with frequent clustering of the same event for women and men. First, Figure \ref{fig:Track_dendrogram} has six instances where the equivalent event for men and women is determined to be most similar (100m, 800m, 1500m, 3K, 5K, and 10K). Furthermore, the dendrogram displays a sub-cluster of great affinity in the bottom left quadrant, composed of all short and middle-distance events of both genders.

Figure \ref{fig:Field_dendrogram} shows the alignment dendrogram between field events. The dendrogram consists of one dominant cluster of primarily throwing events and a smaller second cluster of jumping events. The first major cluster includes both genders' hammer throw, discus, javelin, and shot put events. This major cluster consists of two subclusters, which exhibit more pronounced affinity within themselves. The second cluster consists of the men and women's high jump and men's long jump. Like Figure \ref{fig:Track_dendrogram}, the first difference behaviors are broadly grouped by event; this effect is more pronounced in track than field events. For example, the men and women's 100 and 200m are closely clustered together, which may reflect substantial intersection in the athletes performing in these events.

This finding is by no means inevitable: one could imagine a scenario where the most similar behavior in these first differences is found between events within the same gender, that is, where athletic improvements are most closely associated among athletes competing of the same gender. This is shown not to be the case. Instead, the results in Figure \ref{fig:AlignmentDendrograms} suggest that improvements are passed on rather uniformly between men and women. This could be due to breakthroughs in technology, technique, strength and conditioning, for each event, being passed on to men and women simultaneously.

\section{Trajectory modelling and anomaly identification}
\label{sec:TrajectoryModelling}

Having observed broad similarity between (first differences in) scores between men and women in the same event in Section \ref{sec:AlignmentAnalysis}, we now wish to approach the relationship between male and female scores differently and identify any events where there is relatively less consistency in scores. This section aims to study trends over time in scores, moving away from the first differences of the last section. As before, let $x_i(t), y_i(t), i=1,...,N, t=1,...,T$ be the multivariate time series consisting of the mean of the top $m=100$ women's and men's scores, respectively. While men and women are analyzed in conjunction, we never simultaneously compare track events to field events quantitatively, only descriptively, so our notation suppresses the dependence on these choices. Hence, rather than having $N=16$ total events, our notation denotes $N=8$ events separately for each of the track and field events.

To determine the consistency of evolutionary performance in men and women's athletics events, we normalize each event's time series for an appropriate comparison. To determine each men's normalized trajectory we first compute $\|y_i \| = \sum^{T}_{t=1} y_i(t)$, and then define $\tilde{y}_i(t) = \frac{y_i(t)}{\|y_i \|}$. We similarly define women's normalized trajectories $\tilde{x}_i(t)$. Then, we may compute distances between normalized trajectories by $D^y_{ij} = \sum^{T}_{t=1} |\tilde{y}_i(t) - \tilde{y}_j(t)|$ and $D^x_{ij} = \sum^{T}_{t=1} |\tilde{x}_i(t) - \tilde{x}_j(t)|$, for men and women respectively. We remark that there are alternative ways to normalize such trajectories, such as subtracting the means and dividing by the ranges. As we use the $L^1$ metric to measure distances between the normalized trajectories, it makes consistent sense to use the $L^1$ norm in the normalization.

Next, we apply the linear transformation $A_{ij} = 1 - \frac{D_{ij}}{\max\{D\}}$ to both the men and women's distance matrices, producing affinity matrices $A$ with all entries in $[0,1]$. To determine the consistency in each event's relative trajectory among the separate collections of men and women's athletic events, we compute the \emph{consistency matrix} $C = |A^x - A^y|$. Specifically, $C_{ij}=|A^x_{ij} - A^y_{ij}|.$ Elements in the matrix $C$ close to 0 represent distances between events that are relatively similar among men and women. To determine the most inconsistent events, we compute $a_j = \sum^{N}_{i=1}C_{ij}$ and rank $a_j$, for $j=1,..,N$. Larger values indicate greater inconsistency between the male and female categories of the same events regarding their relationship with other events. We note that these distance, affinity and consistency matrices are computed separately for track and field events, so we suppress the dependence on this choice.

\begin{figure}
    \centering
    \includegraphics[width=0.49\textwidth]{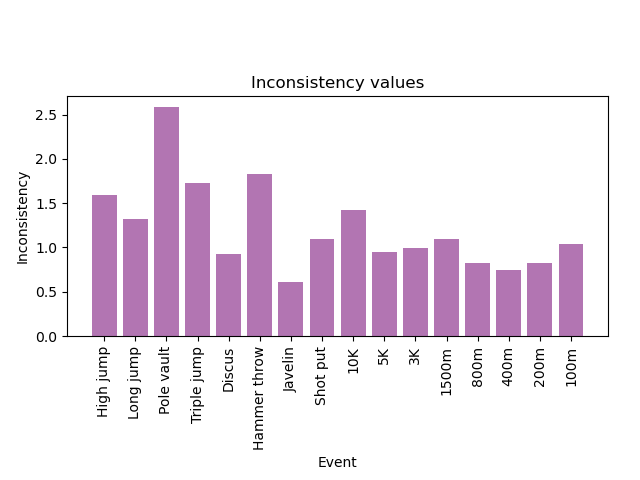}
    \caption{Inconsistency values $a_j$, defined in Section \ref{sec:TrajectoryModelling}, measure extent of inconsistency between men and women's categories of the same event. The pole vault and javelin are the most and least consistent events, as shown in more detail in Figure \ref{fig:NormalizedTrajectories}.}
    \label{fig:TrajectoryAnomalies}
\end{figure}

Figure \ref{fig:TrajectoryAnomalies} documents the inconsistency values $a_j$ of male and female trajectories in each event. The table highlights greater consistency in event trajectories among the track compared to the field events, with median values of 0.97 and 1.46, respectively. Field events display greater variability in their inconsistency values, with a standard deviation of 0.61, compared with 0.21 exhibited by track events. Indeed, both the most and least consistent events are field events, namely the javelin (inconsistency value of 0.61) and the pole vault (value of 2.58).

\begin{figure*}
    \centering
    \begin{subfigure}[b]{0.49\textwidth}
        \includegraphics[width=\textwidth]{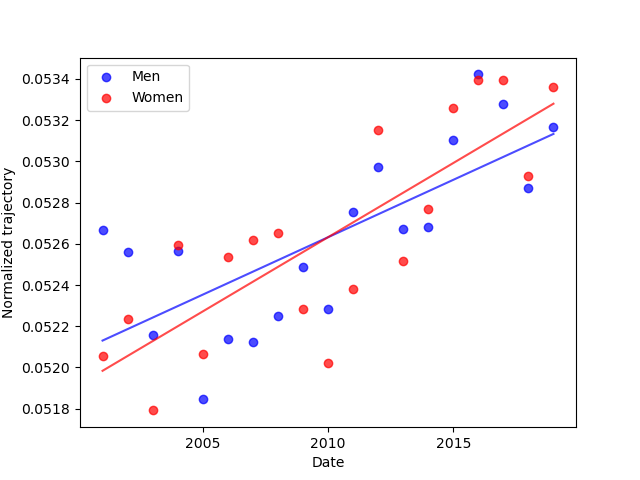}
        \caption{}
        \label{fig:Javelin_consistency}
    \end{subfigure}
    \begin{subfigure}[b]{0.49\textwidth}
        \includegraphics[width=\textwidth]{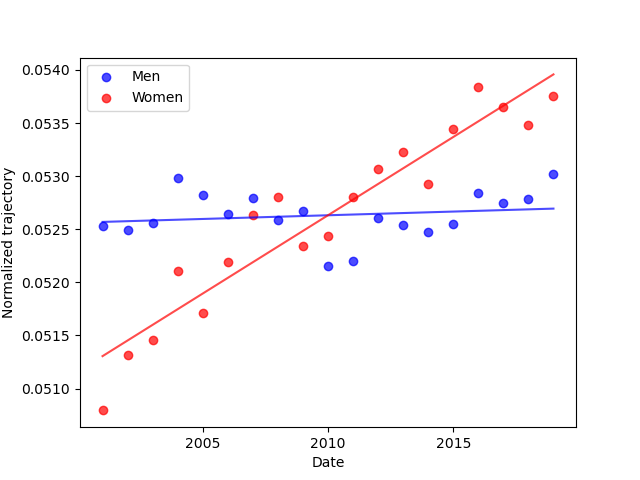}
        \caption{}
        \label{fig:PoleVault_consistency}
    \end{subfigure}
    \caption{Normalized trajectories for (a) the javelin and (b) the pole vault. The javelin displays high consistency in behaviors between men and women, while the pole vault displays low consistency.}
   \label{fig:NormalizedTrajectories}
\end{figure*}

Figures \ref{fig:Javelin_consistency} and \ref{fig:PoleVault_consistency} show the normalized trajectories of athletic scores over time for both men and women in the javelin and pole vault, respectively. It is evident from Figure \ref{fig:NormalizedTrajectories} that the javelin exhibits highly similar trajectories for men and women - both feature a positive linear trend and rather uniform improvement with time. By contrast, the pole vault displays highly varying trends between genders. The women's pole vault shows steady improvement in performance, while the men's event exhibits a predominantly flat trajectory over time.

\section{Geographic concentration analysis}
\label{sec:GeographicConcentration}

Finally, we investigate trends in the composition of the top 100 scoring athletes in each event over time. Our data set records which country each athlete represents in the competitions, and we aim to produce and analyze a time-varying measure of the geographic diversity of athletes. For each event indexed $i$ and time $t=2001,...,2019$ we define an $m \times m$ distance matrix $\Omega^i(t)$ between the top $m=100$ athletes' locations. Specifically, let $\Omega^i(t)_{kl}$ be the geodesic distance between the (centroids of the) countries of nationality of the $k$ and $l$th athletes. For example, if all of the top $m$ athletes came from the same country in a given year, $\Omega^i(t)$ would be an $m \times m$ matrix of zeroes. Larger values throughout the matrix indicate more geographic diversity of athletes, taking geographic locations into account. For an $m \times m$ matrix $A$, we use the Frobenius norm to quantify the total magnitude of the matrix, defined as $\|A\|=( \sum_{k,l=1}^m A^2_{kl})^\frac12$.

For each event $i$, we analyze the following function of time, 
\begin{align}
t \mapsto \| \Omega^i(t) \|,
\end{align}
and compute the total geographic dispersion over time, defined by
\begin{align}
\label{eq:totalgeo}
    G^{i} = \sum^{T}_{t=1} \| \Omega^i(t) \|,
\end{align}
measured in meters.

\begin{figure*}
    \centering
    \begin{subfigure}[b]{0.49\textwidth}
        \includegraphics[width=\textwidth]{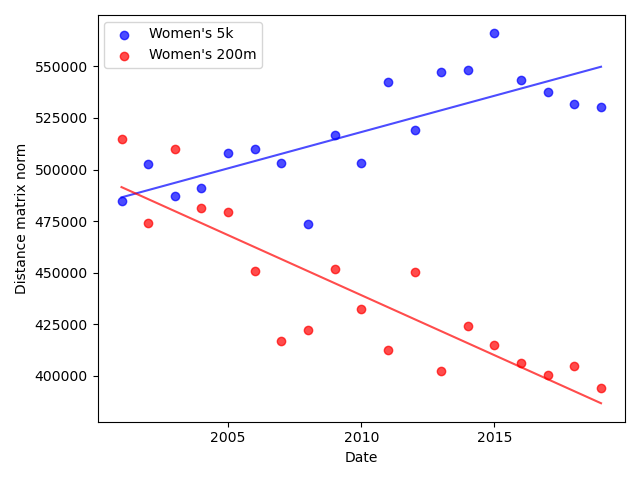}
        \caption{}
        \label{fig:Track_geographic}
    \end{subfigure}
    \begin{subfigure}[b]{0.49\textwidth}
        \includegraphics[width=\textwidth]{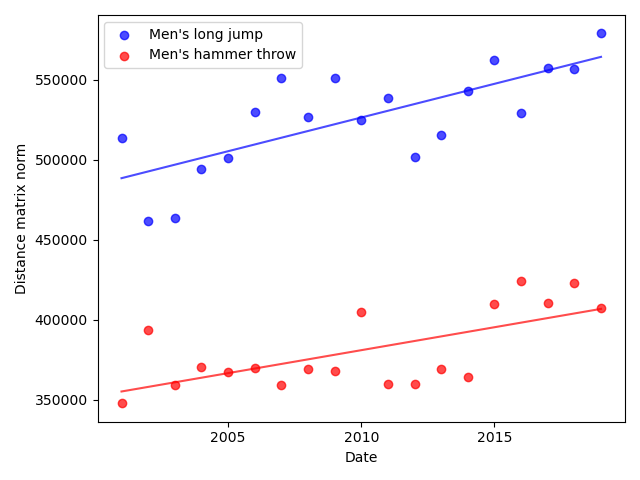}
        \caption{}
        \label{fig:Field_geographic}
    \end{subfigure}
    \caption{Time-varying geographic distance matrix norms for two selected (a) track and (b) field events. Each figure displays the two events with the most and least cumulative geographic dispersion across 2001-2019.}
   \label{fig:GeographicDispersion}
\end{figure*}

Our analysis highlights several trends in the variance of geographic representation among men and women's events. All values of $G^i$, quantifying the total geographic variance for each event, are recorded in Table \ref{tab:GeographicDistanceNorms}. First, a highly similar geographic dispersion is identified between track and field events. The track events produce a mean (and median) value of 9.2 x 10$^6$, while the field events produce a mean (and median) value of 8.8 x 10$^6$. The most geographically concentrated track event is the women's 200m, exhibiting a total value of 8.3 x 10$^6$. The least concentrated track event is the women's 5K, which produces a cumulative value of 9.8 x 10$^6$. Among field events, men's hammer throw consists of the most geographic concentration among the top 100 athletes, while the men's long jump is the most geographically dispersed, with cumulative values of 7.2 x 10$^6$ and 9.9 x 10$^6$, respectively.

Second, a highly similar geographic dispersion is also observed between men and women's events. The mean values among men and women are 9.0 x 10$^6$ and 8.9 x 10$^6$, respectively (with median values both 9.0 x 10$^6$). Among men's events, the most geographic variance in athletes is in the men's long jump, while the greatest concentration is in the men's hammer throw. Among female events, the greatest geographic concentration is in the 5K and 10K, with total values of 9.8 x 10$^6$, while the least geographic dispersion among athletes is in the 200m and high jump, which both have cumulative values of 8.3 x 10$^6$.

Figure \ref{fig:GeographicDispersion} displays the time-varying geographic variance for the aforementioned events with the greatest and least total values. Three of these displayed events exhibit a moderate increase in geographic variance over time, but the women's 200m exhibits a rather pronounced decrease. As a case study, we investigate the changing composition of this event. In Table \ref{tab:200m}, we perform a simple partition of women's 200m athletes in 2001 and 2019 by continent of origin. This contrast confirms Figure \ref{fig:Track_geographic}, showing that the women's 200m has become less geographically diverse over time, primarily due to a large number of athletes from the United States (US), 49 in 2019.

\begin{figure*}
    \centering
    \begin{subfigure}[b]{0.49\textwidth}
        \includegraphics[width=\textwidth]{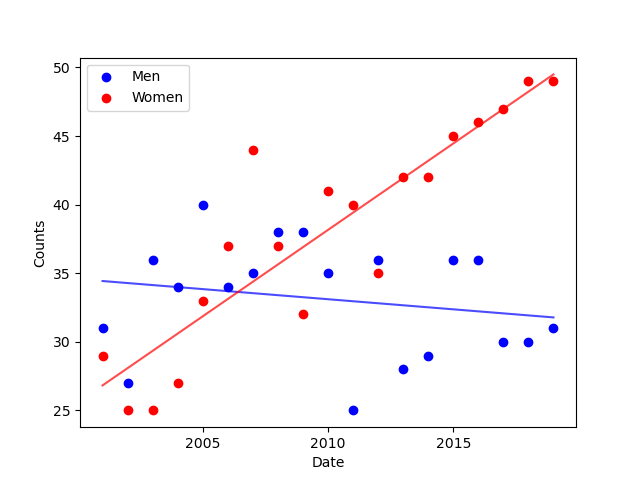}
        \caption{}
        \label{fig:US_200m}
    \end{subfigure}
    \begin{subfigure}[b]{0.49\textwidth}
        \includegraphics[width=\textwidth]{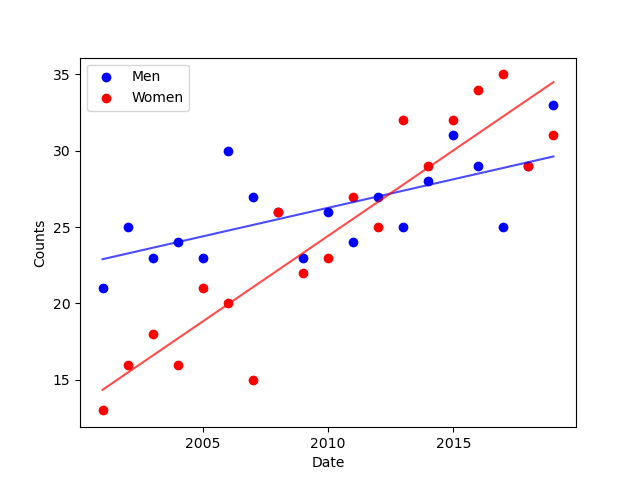}
        \caption{}
        \label{fig:US_shotput}
    \end{subfigure}
    \caption{Number of US athletes featured among the top 100 scorers each year for (a) the 200 meters and (b) the shot put. A considerable increase is observed in the number of US women, while the number of US men is relatively flat.}
   \label{fig:USnumbers}
\end{figure*}

To investigate this further, we plot the number of US athletes among the top 100 as a function of time in Figure \ref{fig:USnumbers}. Figure \ref{fig:200m_women} depicts the counts for US athletes in the men and women's 200m. To include a field event as well, Figure \ref{fig:US_shotput} depicts the analogous counts for the shot put. In each sport, we notice a considerable increase in the proportion of women athletes from the US.

\begin{table}
\begin{center}
\begin{tabular}{ |p{4cm}||p{2.2cm}|p{2.2cm}|}
 \hline
 Continent & 2001 & 2019 \\
 \hline
North America & 32 & 51$^1$\\
Europe & 29 & 19\\
Caribbean/Central America & 14 & 15\\
Africa & 9 & 7\\
Asia & 10 & 4\\
South America & 3 & 6\\
Oceania & 3 & 0\\
\hline
\end{tabular}
\caption{Distribution of women's 200m athletes in 2001 and 2019 by continent of origin, highlighting increased geographic concentration. $^1$In 2019, 49 women were from the US.}
\label{tab:200m}
\end{center}
\end{table}

\begin{table}
\begin{center}
\begin{tabular}{ |p{4cm}||p{2.2cm}|}
 \hline
 \multicolumn{2}{|c|}{Total geographic dispersion} \\
 \hline
 Event & $G^{i}$ \\
 \hline
 Men's high jump & 8.9 x 10$^6$ \\
 Women's high jump & 8.3 x 10$^6$ \\
 Men's long jump & 9.9 x 10$^6$ \\  
 Women's long jump & 9.0 x 10$^6$ \\  
 Men's pole vault & 8.5 x 10$^6$ \\  
 Women's pole vault & 8.9 x 10$^6$ \\  
 Men's triple jump & 9.3 x 10$^6$ \\  
 Women's triple jump & 8.6 x 10$^6$ \\  
 Men's discus & 8.4 x 10$^6$ \\  
 Women's discus & 9.4 x 10$^6$ \\  
 Men's hammer throw & 7.2 x 10$^6$ \\
 Women's hammer throw & 8.3 x 10$^6$ \\  
 Men's javelin & 8.7 x 10$^6$ \\  
 Women's javelin & 8.7 x 10$^6$ \\  
 Men's shot put & 8.8 x 10$^6$ \\  
 Women's shot put & 9.3 x 10$^6$ \\  
  Men's 10K & 9.7 x 10$^6$ \\  
 Women's 10K & 9.8 x 10$^6$ \\  
 Men's 5K & 8.6 x 10$^6$ \\  
 Women's 5K & 9.8 x 10$^6$ \\  
 Men's 3K & 9.0 x 10$^6$\\  
 Women's 3K & 9.2 x 10$^6$\\  
  Men's 1500m & 9.4 x 10$^6$ \\  
 Women's 1500m & 9.0 x 10$^6$\\  
 Men's 800m & 9.4 x 10$^6$ \\  
 Women's 800m & 8.8 x 10$^6$ \\  
 Men's 400m & 9.8 x 10$^6$\\  
 Women's 400m & 9.2 x 10$^6$ \\  
 Men's 200m & 9.2 x 10$^6$\\  
 Women's 200m & 8.3 x 10$^6$ \\  
 Men's 100m & 9.0 x 10$^6$\\  
 Women's 100m & 8.6 x 10$^6$ \\  
\hline
\end{tabular}
\caption{Total geographic dispersion over time $G^i$, defined in (\ref{eq:totalgeo}), for each event. Limited variation is observed between different events.}
\label{tab:GeographicDistanceNorms}
\end{center}
\end{table}

\section{Conclusion}

In this work, we have performed several quantitative analyses to identify patterns in the performance of athletes, the relationship between men and women's trends within the same event, and the geographic composition of athletes over time.

This work takes a different approach to other quantitative analyses focused on modelling sport. Rather than modelling yearly performance data among a very small sample of results \cite{Wu2021_sport,Einmahl2008,Robinson1995,Gembris2002,Baro1999} (usually just the top scorer), this study is interested in broader patterns in performance. Furthermore, many of the techniques used throughout this paper may not provide the same level of insight when small annual samples of results lead to extremely volatile trends. Accordingly, throughout most of our analysis we study the top $m=100$ scorers per event per year.

First, in Section \ref{sec:Olympic effect}, we implemented a regression analysis that revealed linear trends in improving average scores over time. By contrasting the fits of several models, we observed a positive Olympic effect in almost every event, where average scores are superior during Olympic years. While not necessarily surprising, this finding was by no means inevitable, as highly significant athletic events, such as the World Athletic Championships, occur more regularly. Nonetheless, it seems that the Olympic Games produce perhaps the highest levels of training and dedication to achieving the best scores possible. There may be several nuanced reasons for this effect. On the one hand, the increased media attention during the Olympics may increase performance; on the other hand, anxiety and stress may lead to decreased performances.\cite{sportsjournal_anxiety}

Next, we examined the linearity of these fits in more detail in Section \ref{sec:curvature}. Perhaps surprisingly, and contrary to some recent reporting, \cite{Berthelot2010,Guardian_Olympics} we observed no evidence that average performance is ``leveling off'' close to the present day. This finding may differ with a careful analysis of world records, namely the single top athletic performance by year, but no evidence was observed when analyzing the top 100 athletes. We remark that there may be considerable limitations to analyzing just the single top performance. For example, no athlete has so far surpassed Usain Bolt's world-record performance in the men's 100m or 200m events, set in 2009. An incorporation of the top 100 athletes provides a better view of the holistic state of the elite performance in any candidate event. On the other hand, it is conceivable that incorporating more data, either further into the past or into the future could exhibit some leveling off.

We must also note that these two sections carry notable limitations and one should be cautious interpreting these results. For example, non-significant coefficients could be an artefact of the low number of data points (just 19) per event. More broadly, a significant coefficient (such as $\beta_1$ for the year-on-year improvement) is always more interpretable than a non-significant coefficient, which is not conclusive evidence of anything. The fact our methods do not broadly produce evidence of performance levelling off is related both to the choice of method and the question of study. Specifically, our methods (such as regression) work well in the context of studying average performances (mostly over the top $m=100$ athletes). They would not work as well focused specifically on the top scorer ($m=1$). For this latter topic, many other papers have used quite different statistical techniques.\cite{Wu2021_sport,Einmahl2008,Robinson1995,Gembris2002,Baro1999}

Continuing our study of the first differences, we observed substantial alignment between the men and women's categories of each event in Section \ref{sec:AlignmentAnalysis}. Even beyond the Olympic effect, men and women's average scores moved in the same direction on a yearly basis to a considerable extent. Clustering based on our normalized inner products frequently paired men and women's categories of the same events together. This suggests that sports may be characterized by improvements in technique and training that benefit the men and women's competitions simultaneously. Interestingly, short and middle distance running exhibited pronounced similarity which may reflect the high degree of intersection in athlete registration among these events. For instance, it is not unusual for the same athlete to represent their country in the 100m and 200m or 200m and 400m. Usain Bolt of Jamaica and Michael Johnson of the US are two examples of such an overlap.

In Section \ref{sec:TrajectoryModelling}, we took a different approach, this time seeking anomalies in the relationship between men and women's categories of the same event. For this purpose, we investigated consistency in the normalized trajectories of scores with time. We were able to identify the pole vault and javelin as the events with the greatest and least consistency in the trajectories of men and women's events. We failed to identify higher levels of structure among our trajectory anomalies.

Finally, Section \ref{sec:GeographicConcentration} turned our attention to the athletes themselves, specifically their countries of origin and the changing composition of events over time. Here, we obtained a surprising result: a high degree of similarity in geographic variance across all studied events, track and field, men and women. This may counter a popular conception that certain events are more dominated by smaller selections of countries, such as sprint events by athletes from the US and Jamaica.\cite{Guardian_Jamaica} Instead, with some exceptions, almost every event exhibited proportionate variance of athletes from across the world. Some trends in changing levels of geographic variance over time were observed, particularly for some women's events. Future work can study the proportion of athletes from the US and investigate whether this proportion is increasing, especially in women's events.

Overall, this work introduces methods to study the evolution of elite athletes' performance over the past two decades. We identify structural similarities and dissimilarities among various track and field events and the performance of men and women. In addition, we highlight surprising homogeneity in time-varying geographic composition among various events' athletic representation. Although these methods were developed for this application specifically, we believe that our techniques could be applied beyond athletics and more broadly beyond competitive sport. Researchers interested in identifying structure in any phenomenon evolving over time and comparing the evolutionary performance of subgroups could apply these techniques effectively.

\section*{Data availability}
The data that support the findings of this study are openly available at Ref. \onlinecite{olympicsdata}.

\begin{acknowledgments}
The authors would like to thank Kirsten Doert Eccles from the Melbourne Centre for Data Science, who first proposed the idea on writing a paper studying human performance and the Olympics. The authors would also like to thank Nathaniel Bloomfield for his help in accessing the data.

\end{acknowledgments}

\appendix

\section{Supplementary experiments on the linear regression}
\label{app:regression}

\begin{figure*}
    \centering
    \begin{subfigure}[b]{0.33\textwidth}
        \includegraphics[width=\textwidth]{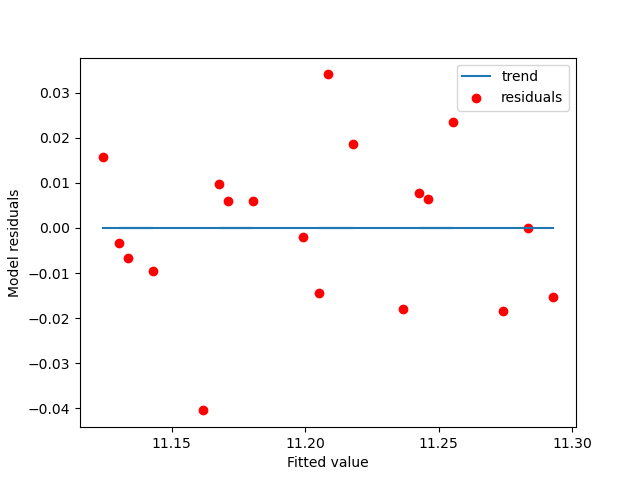}
        \caption{}
        \label{fig:residuals1}
    \end{subfigure}
    \begin{subfigure}[b]{0.33\textwidth}
        \includegraphics[width=\textwidth]{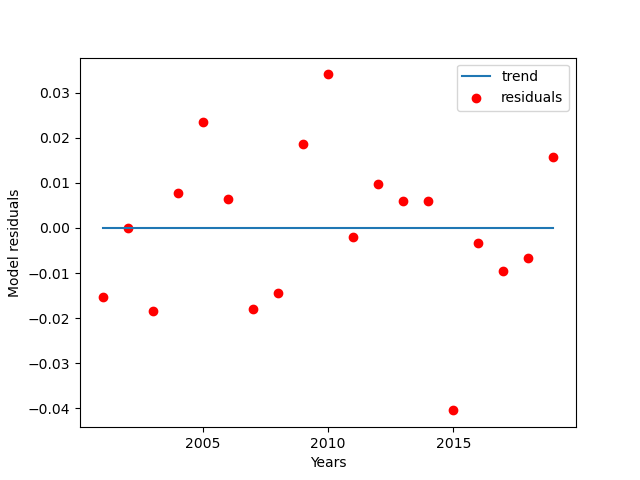}
        \caption{}
        \label{fig:residuals2}
    \end{subfigure}
    \begin{subfigure}[b]{0.33\textwidth}
        \includegraphics[width=\textwidth]{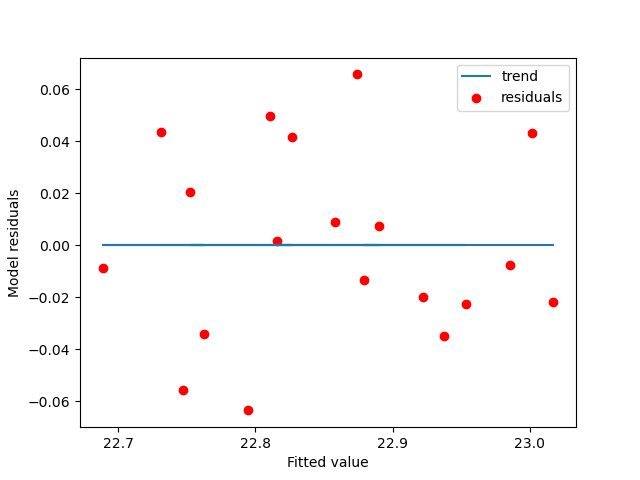}
        \caption{}
        \label{fig:residuals3}
    \end{subfigure}
    \begin{subfigure}[b]{0.33\textwidth}
        \includegraphics[width=\textwidth]{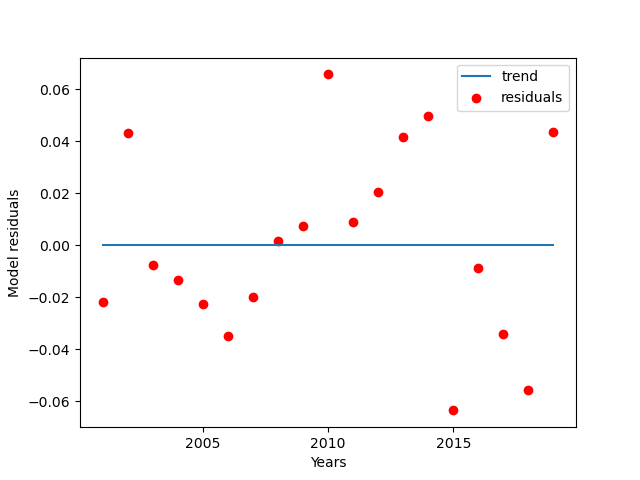}
        \caption{}
        \label{fig:residuals4}
    \end{subfigure}
    \begin{subfigure}[b]{0.33\textwidth}
        \includegraphics[width=\textwidth]{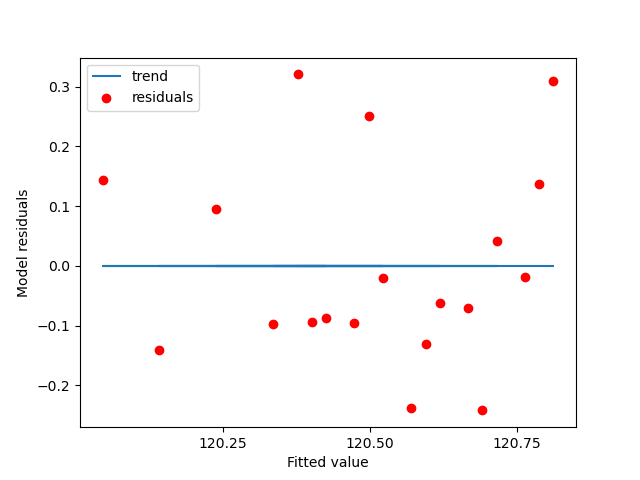}
        \caption{}
        \label{fig:residuals5}
    \end{subfigure}
    \begin{subfigure}[b]{0.33\textwidth}
        \includegraphics[width=\textwidth]{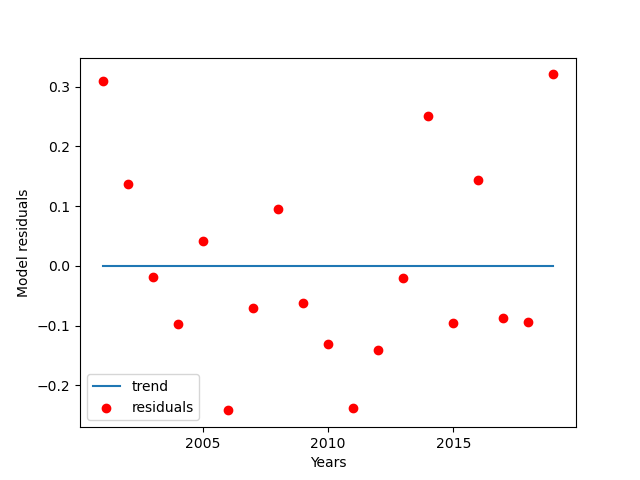}
        \caption{}
        \label{fig:residuals6}
    \end{subfigure}
    \begin{subfigure}[b]{0.33\textwidth}
        \includegraphics[width=\textwidth]{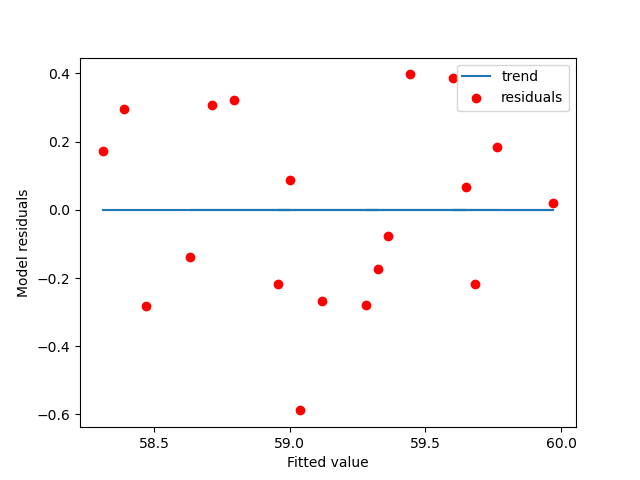}
        \caption{}
        \label{fig:residuals7}
    \end{subfigure}
    \begin{subfigure}[b]{0.33\textwidth}
        \includegraphics[width=\textwidth]{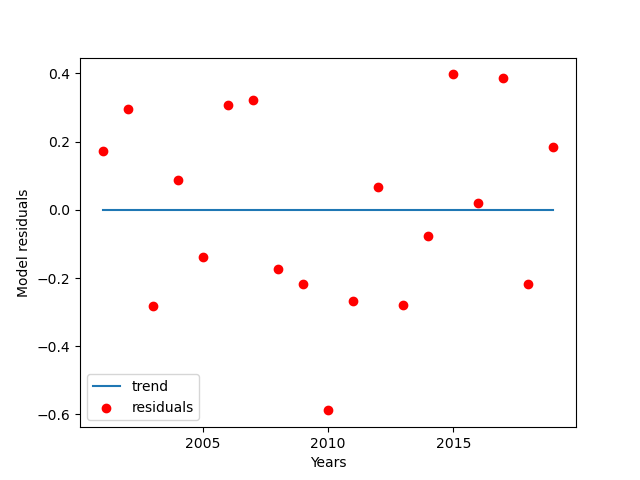}
        \caption{}
        \label{fig:residuals8}
    \end{subfigure}
    \begin{subfigure}[b]{0.33\textwidth}
        \includegraphics[width=\textwidth]{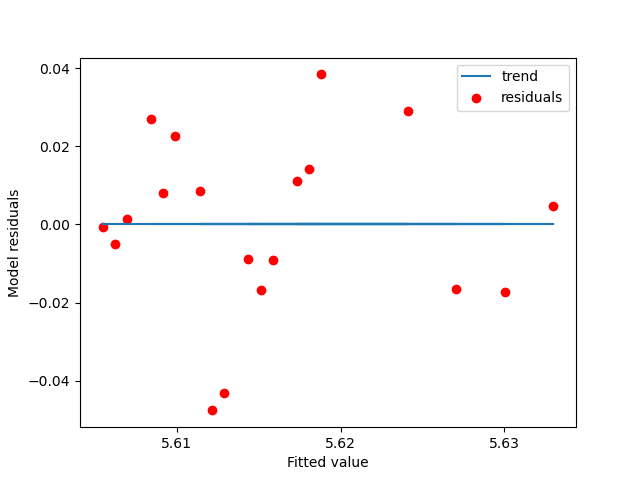}
        \caption{}
        \label{fig:residuals9}
    \end{subfigure}
    \begin{subfigure}[b]{0.33\textwidth}
        \includegraphics[width=\textwidth]{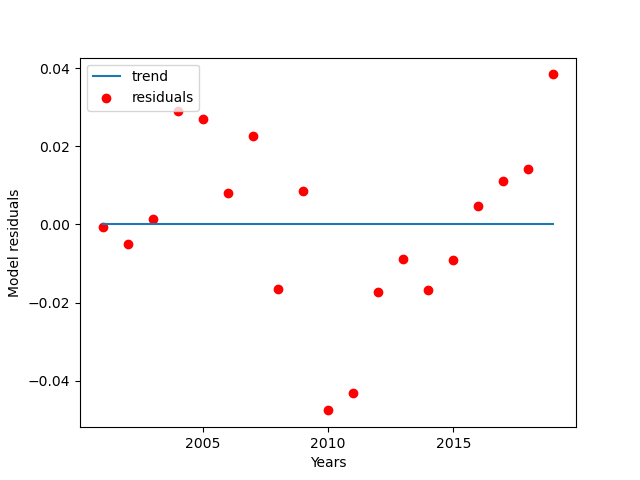}
        \caption{}
        \label{fig:residuals10}
    \end{subfigure}
    \caption{In the above figures, we display model residuals vs fitted values and years (the time covariate). In (a,b) we plot residuals for the 100 metres women in (c,d) for the 200 metres women, (e,f) for 800 m women, (g,h) women's javelin, (i,j) men's pole vault.}
   \label{fig:residuals}
\end{figure*}

In Figure \ref{fig:residuals}, we demonstrate the homoscedasticity of residuals for a collection of events, complementing the linear regressions performed in Section \ref{sec:Olympic effect}. The plots show in most cases that residuals are satisfactorily distributed with mean zero and rather constant variance.

In Tables \ref{tab:Model1AICBIC}, \ref{tab:Model2AICBIC} and \ref{tab:Model3AICBIC}, respectively, we document Akaike information criteria (AIC) and Bayesian information criteria (BIC) scores \cite{Vrieze2012} for each of Models 1,2 and 3 from Section \ref{sec:Olympic effect}.

\begin{table*}
\begin{center}
\begin{tabular}{ |p{3.5cm}||p{2.2cm}|p{2.2cm}|p{2.2cm}|p{2.2cm}|}
 \hline
 \multicolumn{5}{|c|}{Model 1 AIC and BIC} \\
 \hline
 Event & $m=100$ AIC & $m=10$ AIC  & $m=100$ BIC & $m=10$ BIC \\
 \hline
 Men's high jump & -147.2 & -113.8 & -145.3 & -111.9 \\
 Women's high jump & -145.5 & -116.6 & -143.6 & -114.7 \\
 Men's long jump & -94.2 & -66.6 & -92.4 & -64.7 \\
 Women's long jump & -81.3 & -59.5 & -79.4 & -57.6 \\
 Men's pole vault & -85.4 & -77.7 & -83.5 & -75.8 \\
 Women's pole vault & -86.3 & -66.9 & -84.4 & -65.0 \\
 Men's triple jump & -50.9 & -46.2 & -49.0 & -44.4 \\
 Women's triple jump & -47.0 & -26.5 & -45.1 & -24.6 \\
 Men's discus & 14.2 & 34.2 & 16.1 & 36.0 \\
 Women's discus & 24.7 & 37.0 & 26.6 & 38.9 \\
 Men's hammer throw & 19.5 & 26.8 & 21.4 & 28.7 \\
 Women's hammer throw & 50.8 & 60.4 & 52.7 & 62.3 \\
 Men's javelin & 27.3 & 58.7 & 29.2 & 60.6 \\
 Women's javelin & 15.4 & 26.1 & 17.3 & 27.9 \\
 Men's shot put & -25.8 & -7.5 & -23.9 & -5.6 \\
 Women's shot put & -24.3 & -13.8 & -22.5 & -11.9 \\
  Men's 10K & 132.8 & 142.8 & 134.7 & 144.7 \\
 Women's 10K & 149.1 & 176.4 & 150.1 & 178.2 \\
 Men's 5K & 85.6 & 98.7 & 87.5 & 100.6 \\
 Women's 5K & 107 & 104.5 & 108.9 & 106.3 \\
 Men's 3K & 75.4 & 83.6 & 77.3 & 85.5 \\
 Women's 3K & 101.1 & 124.5 & 103.0 & 126.4 \\
  Men's 1500m & 19.4 & 42.0 & 21.3 & 43.9 \\
 Women's 1500m & 29.0 & 57.7 & 30.1 & 59.6 \\
 Men's 800m & -21.5 & 15.9 & -19.6 & 17.8 \\
 Women's 800m & 2.4 & 30.5 & 4.3 & 32.4 \\
 Men's 400m & -38.1 & -14.2 & -36.2 & -12.3 \\
 Women's 400m & -19.7 & -10.8 & -17.8 & -8.9 \\
 Men's 200m & -74.3 & -51.3 & -72.4 & -49.4 \\
 Women's 200m & -55.1 & -37.4 & -53.2 & -35.5 \\
 Men's 100m & -99.0 & -68.0 & -97.1 & -66.1 \\
 Women's 100m & -92.2 & -65.7 & -90.3 & -63.8 \\
\hline
\end{tabular}
\caption{AIC and BIC values for Model 1 regression fits, for both $m=100$ and $m=10$.}
\label{tab:Model1AICBIC}
\end{center}
\end{table*}

\begin{table*}
\begin{center}
\begin{tabular}{ |p{3.5cm}||p{2.2cm}|p{2.2cm}|p{2.2cm}|p{2.2cm}|}
 \hline
 \multicolumn{5}{|c|}{Model 2 AIC and BIC} \\
 \hline
 Event & $m=100$ AIC & $m=10$ AIC  & $m=100$ BIC & $m=10$ BIC \\
 \hline
 Men's high jump & -148.8 & -113.3 & -145.9 & -110.5 \\
 Women's high jump & -147.2 & -116.3 & -144.4 & -113.5 \\
 Men's long jump & -94.7 & -65.0 & -91.9 & -62.1 \\
 Women's long jump & -86.8 & -68.1 & -83.9 & -65.3 \\
 Men's pole vault & -85.1 & -75.8 & -82.3 & -72.9 \\
 Women's pole vault & -93.8 & -70.0 & -90.9 & -67.0 \\
 Men's triple jump & -53.7 & -44.5 & -50.1 & -41.7 \\
 Women's triple jump & -51.7 & -33.7 & -48.8 & -30.9 \\
 Men's discus & -2.8 & 34.9 & 0.07 & 37.7 \\
 Women's discus & 12.3 & 37.7 & 15.1 & 40.5 \\
 Men's hammer throw & 5.4 & 26.0 & 8.2 & 28.8 \\
 Women's hammer throw & 45.4 & 61.3 & 48.2 & 64.1 \\
 Men's javelin & 27.5 & 59.3 & 30.3 & 62.2 \\
 Women's javelin & 10.3 & 27.0 & 13.1 & 29.9 \\
 Men's shot put & -26.2 & -6.1 & -23.4 & -3.3 \\
 Women's shot put & -27.6 & -22.2 & -24.7 & -19.4 \\
  Men's 10K & 127.4 & 143.5 & 130.3 & 146.3 \\
 Women's 10K & 145.4 & 173.3 & 148.2 & 176.1 \\
 Men's 5K & 84.8 & 100.4 & 87.7 & 103.2 \\
 Women's 5K & 104.7 & 106 & 107.5 & 108.8 \\
 Men's 3K & 75.0 & 85.5 & 77.9 & 88.4 \\
 Women's 3K & 102.1 & 124.9 & 104.9 & 127.7 \\
  Men's 1500m & 19.8 & 43.5 & 22.7 & 46.4 \\
 Women's 1500m & 29.1 & 59.7 & 31.9 & 62.6 \\
 Men's 800m & -24.5 & 16.2 & -21.7 & 19.1 \\
 Women's 800m & -8.9 & 26.2 & -6.1 & 29.1 \\
 Men's 400m & -36.3 & -12.5 & -33.5 & -9.7 \\
 Women's 400m & -24.8 & -10.6 & -22 & -7.7 \\
 Men's 200m & -73.1 & -49.7 & -70.2 & -46.8 \\
 Women's 200m & -66.9 & -42.3 & -64.0 & -39.4 \\
 Men's 100m & -103.8 & -69.4 & -101 & -66.6 \\
 Women's 100m & -94.9 & -71.3 & -92.1 & -68.4 \\
\hline
\end{tabular}
\caption{AIC and BIC values for Model 2 regression fits, for both $m=100$ and $m=10$.}
\label{tab:Model2AICBIC}
\end{center}
\end{table*}

\begin{table*}
\begin{center}
\begin{tabular}{ |p{3.5cm}||p{2.2cm}|p{2.2cm}|p{2.2cm}|p{2.2cm}|}
 \hline
 \multicolumn{5}{|c|}{Model 3 AIC and BIC} \\
 \hline
 Event & $m=100$ AIC & $m=10$ AIC  & $m=100$ BIC & $m=10$ BIC \\
 \hline
 Men's high jump & -145.4 & -109.5 & -140.7 & -104.8 \\
 Women's high jump & -152.1 & -118.3 & 147.4 & 113.6 \\
 Men's long jump & -92.5 & -61.0 & -87.8 & -56.3 \\
 Women's long jump & -91.3 & -66.6 & -86.6 & -61.9 \\
 Men's pole vault & -82.7 & -73.8 & -78.0 & -69.1 \\
 Women's pole vault & -92.8 & -68.8 & -88.1 & -64.1 \\
 Men's triple jump & -51.3 & -44.7 & -46.6 & -40.0 \\
 Women's triple jump & -51.2 & -34.6 & -46.4 & -29.9 \\
 Men's discus & 0.70 & 37.5 & 5.4 & 42.3 \\
 Women's discus & 14.4 & 41.5 & 19.1 & 46.2 \\
 Men's hammer throw & 1.1 & 25.5 & 5.8 & 30.2 \\
 Women's hammer throw & 47.2 & 65.2 & 52.0 & 69.9 \\
 Men's javelin & 30.8 & 63.3 & 35.6 & 68.0 \\
 Women's javelin & 13.7 & 27.5 & 18.4 & 32.2 \\
 Men's shot put & -22.9 & -3.8 & -18.2 & .98 \\
 Women's shot put & -24.0 & -18.4 & -19.3 & -13.7 \\
  Men's 10K & 117.0 & 139.7 & 121.7 & 144.4 \\
 Women's 10K & 143.9 & 173.8 & 148.6 & 178.5 \\
 Men's 5K & 88.7 & 103 & 93.4 & 107.7 \\
 Women's 5K & 106.6 & 108.7 & 111.3 & 113.4 \\
 Men's 3K & 74.6 & 87.3 & 79.4 & 92.0 \\
 Women's 3K & 99.2 & 125.9 & 103.9 & 130.6 \\
  Men's 1500m & 18.8 & 47.2 & 23.6 & 51.9 \\
 Women's 1500m & 33.0 & 63.2 & 37.7 & 67.9 \\
 Men's 800m & -21.6 & 17.4 & -16.9 & 22.1 \\
 Women's 800m & -5.3 & 29.3 & -.55 & 34.0 \\
 Men's 400m & -33.6 & -9.6 & -28.9 & -4.9 \\
 Women's 400m & -21.5 & -8.5 & -16.8 & -3.7 \\
 Men's 200m & -70.6 & -50.3 & -65.9 & -45.6 \\
 Women's 200m & -64.0 & -39.3 & -59.3 & -34.6 \\
 Men's 100m & -103.4 & -66.0 & -98.7 & -61.2 \\
 Women's 100m & -95.8 & -76.0 & -91.1 & -71.2 \\
\hline
\end{tabular}
\caption{AIC and BIC values for Model 3 regression fits, for both $m=100$ and $m=10$.}
\label{tab:Model3AICBIC}
\end{center}
\end{table*}

\section{Geographic Variance Gini}

\begin{figure*}
    \centering
    \begin{subfigure}[b]{0.49\textwidth}
        \includegraphics[width=\textwidth]{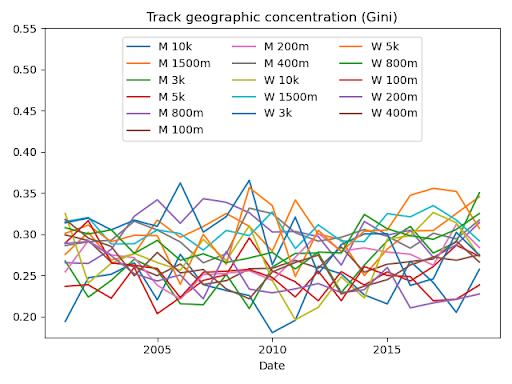}
        \caption{}
        \label{fig:trackgini100}
    \end{subfigure}
    \begin{subfigure}[b]{0.49\textwidth}
        \includegraphics[width=\textwidth]{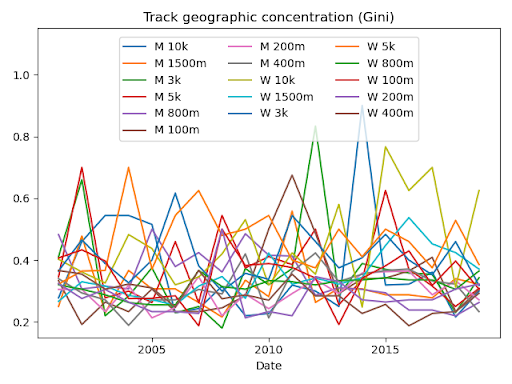}
        \caption{}
        \label{fig:trackgini10}
    \end{subfigure}
\begin{subfigure}[b]{0.49\textwidth}
        \includegraphics[width=\textwidth]{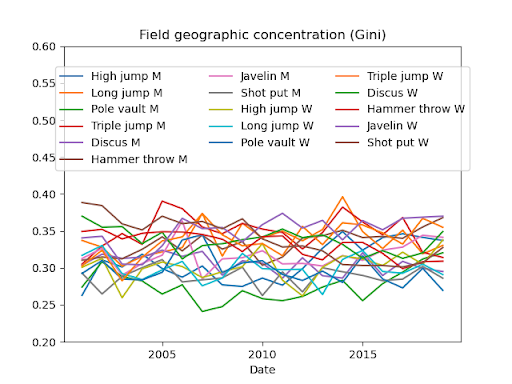}
        \caption{}
        \label{fig:fieldgini100}
    \end{subfigure}
\begin{subfigure}[b]{0.49\textwidth}
        \includegraphics[width=\textwidth]{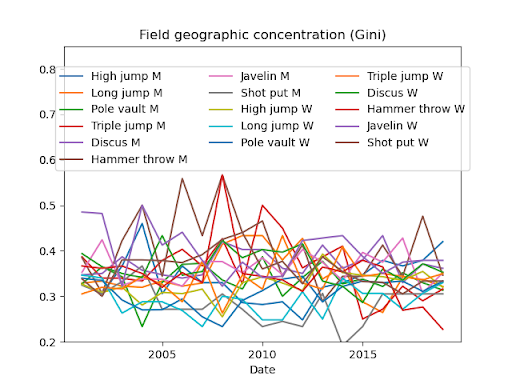}
        \caption{}
        \label{fig:fieldgini10}
    \end{subfigure}
    \caption{Gini coefficients computed directly from the list of countries of athletes' nationalities for each event on a year-by-year basis. We display Gini coefficients for (a) top 100 scorers in track events each year (b) top 10 scorers in track events (c) top 100 scores in field events (d) top 10 scorers in field events.}
   \label{fig:giniscores}
\end{figure*}

In Figure \ref{fig:giniscores}, we complement our study of geographic variation in Section \ref{sec:GeographicConcentration} with a more straightforward measure of athletes' diversity of nationalities each year. We display computed Gini coefficients for track and field events, both according to the top 100 and 10 scorers each year. Compared to Section \ref{sec:GeographicConcentration}, we see slightly more variability between different events. These different results can be explained by the different method of calculation, which does not take into account geographic dispersion.

\bibliography{__newreferences}
\end{document}